\documentstyle[eqsecnum,aps,epsfig]{revtex}
\newcommand{\nn}{\nonumber}
\newcommand{\be}{\begin{equation}}
\newcommand{\ee}{\end{equation}}
\newcommand{\bd}{\begin{displaymath}}
\newcommand{\ed}{\end{displaymath}}
\newcommand{\bea}{\begin{eqnarray}}
\newcommand{\eea}{\end{eqnarray}}
\renewcommand{\paragraph}[1]{
\vspace{.8mm}\par\noindent {\sl #1}\\
\vspace{0.2mm} }

\def\Tr{{\rm Tr}\ }

\def\ket#1{\left|#1\right>}
\def\bra#1{\left<#1\right|}

\def\a{\alpha}
\def\b{\beta}
\def\g{\gamma}

\begin{document}
\draft
\preprint{
SU-ITP/01-27\\
hep-th/0106030}
\title{Why Matrix theory works for oddly shaped membranes}
\author{Yonatan Zunger}
\address{Department of Physics\\Stanford University\\Stanford, CA 94305-4060}
\date{\today}
\maketitle
\begin{abstract}
We give a simple proof of why there is a Matrix theory approximation for a membrane
shaped like an arbitrary Riemann surface. As corollaries, we show that noncompact
membranes cannot be approximated by matrices and that the Poisson algebra on
any compact phase space is $U(\infty)$. The matrix approximation does not appear to
work properly in theories such as IIB string theory or bosonic membrane theory where
there is no conserved 3-form charge to which the membranes couple.
\end{abstract}

\section{Matrix theory and Poisson brackets}

The idea of approximating various aspects of stringy dynamics, especially the dynamics
of membranes, by Matrix theories -- quantum field theories in low (0+1 and 1+1) dimensions
whose fields are matrix-valued -- has been around for some time. In the early 1980's,
Goldstone \cite{Goldstone} and Hoppe \cite{Hoppe} independently examined the
action for a relativistic membrane and showed that, for
membranes of spherical geometry, the Nambu-Goto action can be approximated to arbitrary
precision by the dimensional reduction of super-Yang-Mills theory to 0+1 dimensions,
where the gauge group is $SU(N)$. As $N\rightarrow\infty$, this gauge group
converged to the group of area-preserving diffeomorphism of the sphere, which is a symmetry
of the original action. This action was the first and simplest example of a matrix model,
and sparked the study of matrix membranes. Some of the important developments
in this field since then are the extension of this result to the supersymmetric case,
\cite{deWit} the Matrix
theory of toroidal membranes, \cite{MatrTorA,MatrTorB,MatrTorC,MatrTorD,MatrTorE} and the
understanding, at least in theory, of how
to do Matrix theory for other topologies of membrane. \cite{OtherPf} A recent comprehensive
review has been given by Taylor. \cite{Taylor}

The matrix approximation was immediately realized to be valuable both as a way for understanding
the membranes which appeared in supergravity (and later, of course, in string theory) and
as a method of regularizing field theories which is compatible with supersymmetry.
Today Matrix theory (in its more sophisticated forms, relating membranes to systems of
D0-branes) is one of our few nonperturbative handles on
M-theory. But as we shall see below, there is a key step in the matrix membrane argument
having to do with the nature of Poisson brackets on membranes which has been implicitly
asserted but not yet proven. In this paper we will prove this conjecture, namely that the
Poisson brackets on any Riemann surface form the algebra $U(\infty)$, and thus put the matrix
membrane on a more solid footing.
Phrased another way, we will show that the group of area-preserving diffeomorphisms of any
Riemann surface is $SU(\infty)$; these two groups differ by a $U(1)$ factor generated by
the constant functions, which generate the trivial diffeomorphism.\footnote{Infinitesimal
area-preserving diffeomorphisms are given by $x^i\rightarrow x^i+\left\{x^i, \lambda\right\}$,
where the braces are Poisson brackets and $\lambda$ is an arbitrary function. This follows
naturally from the Liouville theorem or from demanding that the Jacobian of an arbitrary
transformation be 1. Constant functions $\lambda$ have vanishing Poisson brackets
with all functions and so do not generate a diffeomorphism.} Phrased another way still, the group of
Hamiltonian flows of a classical system with compact phase space is always
$U(\infty)$.\footnote{Whenever one refers to $U(\infty)$, one must deal with the subtle
issue of to which $U(\infty)$ one is referring; for example, one could define it as
$\bigcup_n U(n)$, or as the set of limits of sequences of unitary operators of increasing
dimension, which is the set of compact unitary operators on a Hilbert space. These groups
have the same Lie algebra but very different topologies. The group to which
we shall refer as $U(\infty)$ in this paper is the latter group, which can also
be described as the group generated by infinite-dimensional clock and shift
matrices. Hoppe and Schaller have given a useful discussion of how this group
is indeed the group of area-preserving diffeomorphisms on the torus, as well
as of the analogous groups for the noncommutative torus. \cite{HopSchal}
Harvey has recently summarized the difference between the various definitions
of $U(\infty)$ and discussed these differences in the context of string
theory. \cite{Harvey}}

Let us begin by demonstrating the subtlety in the usual derivation of the matrix membrane.
The action for a supersymmetric membrane in type IIA string theory in the light-cone frame
is \cite{Halpern,BST}
\be
S = \int d^3\xi \ \frac{1}{2}\dot{X}^2 + \bar\theta\g_-\dot\theta -
\frac{1}{4}\left\{X^i, X^j\right\}^2 + \bar\theta\g_-\g_i\left\{X^i,\theta\right\}
\label{eq:superbrane1}
\ee
where the brackets represent Poisson brackets with respect to the membrane world-volume
coordinates. The shape of the membrane (say, a sphere) is encoded by a vacuum
expectation value (VEV) of the $X^i$. The observation which led to matrix membrane theory was
the demonstration \cite{Goldstone,Hoppe} that the Lie algebra of infinitesimal area-preserving
diffeomorphisms on the sphere was none other than $SU(\infty)$, and therefore it could be
approximated at arbitrary order by the matrices of $SU(N)$. To do this, one replaces the
functions $X^i(\xi)$ and $\theta(\xi)$ with $SU(N)$--valued functions of time,
the Poisson brackets with
commutators, and the integral over the space coordinates with a trace in the matrix
space. If one does this, one finds the action
\be
S = \int dt \ \Tr\left[\frac{1}{2}\dot{X}^2 + \bar\theta\g_i\dot\theta -
\frac{1}{4}\left[X^i, X^j\right]^2 + \bar\theta\g_-\g_i\left[X^i, \theta\right]
\right]\ .
\label{eq:superbrane2}
\ee
This action is immediately recognizable as the action for super-Yang-Mills theory
of the gauge group $SU(N)$ dimensionally reduced to 0+1 dimensions in temporal gauge $X^0=0$.
The field strength in this gauge is
\bea
F^{0i} &=& \dot{X}^i \nn \\
F^{ij} &=& -i[X^i, X^j]
\label{eq:fldstr}
\eea
and the action is
\be
S = \int dt\ \Tr \left[\frac{1}{4}F^2 + \bar\theta\g_-\g\cdot\nabla\theta\right]\ ,
\label{eq:superbrane3}
\ee
 This action is also the low-energy action for a system
of $N$ D0-branes, where the eigenvalues of the VEV of the $X^i$ give the positions
of the branes, and the off-diagonal elements represent amplitudes for strings to be
stretched between a given pair of branes.

The problem with this argument is that it relies on the Poisson algebra of the
membrane being $U(\infty)$, independent of the fields. However, the spherical geometry is
encoded only by the values of the $X^i$; the most general set of fluctuations should
include higher-genus surfaces. If the Poisson algebra were to depend on the topology of the
space (and there is no {\em a priori} reason to believe that this should not be the case)
then in the action (\ref{eq:superbrane2}) the gauge group would be a function of the gauge
fields, and so this would not be a Yang-Mills action.
 One cannot arbitrarily restrict the fluctuations to those which
preserve the Poisson algebra, since any such truncation would clearly violate unitarity;
all membrane geometries might appear as intermediate states of the theory.

It is therefore crucial to understand the dependence of the Poisson algebra of a
manifold on its geometry. Our key result is that the Poisson algebra of any Riemann
surface (a connected, compact and orientable surface) is $U(\infty)$. This statement
has been long conjectured but no complete proof has appeared in the literature.\footnote{It
has already been shown that every membrane geometry has a matrix approximation,
but the independence of this approximation on the geometry in the large-$N$ limit remains an
open question. \cite{OtherPf}} Using
this theorem we can close the gap in the preceding argument, since now all finite-energy
fluctuations about a compact background geometry should preserve compactness
and thus leave the Poisson algebra fixed.

Before proving this result, let us consider the meaning of configurations which
violate its three conditions. For disconnected manifolds, we will recover the usual block-diagonal
behavior associated
with disconnected membranes. The Poisson algebra is $U(\infty)^n$, with a factor for
each connected component. This $U(\infty)^n$ is contained in $U(\infty)$, with the
off-diagonal terms becoming light when two membranes are close to each other.

For noncompact manifolds, we will see that
the algebra has continuously infinite dimension (this follows from the fact that one
needs a continuous rather than countable basis to specify functions on a noncompact
surface) and therefore no matrix approximation exists. The simplest example of this is
the matrix plane, where the Poisson bracket $\left\{x,y\right\}=1$ clearly has no
finite-dimensional approximations. Physically this is because a finite number of
D0-branes cannot approximate an infinite surface area. In general, when one attempts
to do this one can recover matrix approximations which are good for certain blocks of the matrix
(those corresponding to the deep interior of the membrane) but the matrices acquire
large elements at their edges. However, it appears to be possible to approximate
these spaces using Matrix theories of spaces with boundary; this involves the
introduction of boundary fields and is substantially more complicated. \cite{Polychr}

Nonorientable membranes such as Klein bottles cannot appear in M-theory or in type
IIA theory since those membranes carry a local charge which is essentially an
orientation on the world-volume. (That is, each small patch of membrane has a
definite sense, and so small patches cannot be consistently glued into unorientable
membrane configurations) They can appear in theories where the membranes are unstable, such
as type IIB theory or bosonic membrane theory. In these theories, we need to be able
to consistently treat such membranes since there is no {\em a priori} obstruction
to their being in the spectrum. However, it is not clear
whether these states can be adequately described by the usual matrix construction.
The problem is that the Poisson brackets do not form a Lie algebra when the membrane
is nonorientable. (This is because the criterion for them to form an algebra is that
the symplectic form be a closed, nondegenerate 2-form. On a 2-manifold every 2-form is
closed, but the existence of a nondegenerate 2-form is precisely the criterion for
orientability) This means that the action cannot be written in the form (\ref{eq:superbrane1})
and so the meaning of the matrix approximation is unclear. It is possible that this
approximation simply misses all nonorientable states; in that case there may be
problems at loop level since these states should emerge as intermediate states. We
therefore restrict our considerations to theories such as type IIA theory or M-theory
where these states do not appear.

\section{Why the Poisson algebra is always $U(\infty)$}

We would now like to prove our main statement:

\medskip

{\bf Theorem.} The Poisson algebra of any compact, connected and
orientable 2-manifold is $U(\infty)$.

\medskip

We will prove this in two parts. Our strategy will be to define a special
class of Lie algebras, the pseudocompact algebras, and show that the only
simple pseudocompact algebra is $SU(\infty)$. We will then show that the Poisson
algebra is $U(1)$ times a simple pseudocompact algebra. The first part of
this proof is unavoidably slightly technical, but it is actually not as
complicated as it may seem. We will show that one can use Dynkin diagrams to
classify pseudocompact algebras in the same way that they classify compact algebras.
The argument for this is very similar to the ordinary classification of finite-dimensional
algebras. The only differences are (1) that a certain technical step in the
proof that the simple roots are enough to reconstruct the algebra does not work
in general for infinite-dimensional algebras, and we must show that it does indeed
work for the pseudocompact case; (2) that all our Dynkin diagrams have infinitely
many circles, and so the actual list of allowable diagrams is slightly different;
and (3) that at the end, we will find four allowable Dynkin diagrams (corresponding
to the four infinite classical algebras) which nontrivially correspond to the same
algebra. Apart from these few points, one can read this
proof with a group theory book in hand and watch an almost exactly identical
development take place. The notation below has been chosen to be compatible with that
found in Georgi \cite{Georgi} since that source is likely to be familiar to
physicists. Some more technical aspects of the proof are closely related to the more
generalized arguments found in Knapp. \cite{Knapp}

\medskip

We begin with a definition: A pseudocompact Lie algebra is a Lie algebra whose
Cartan-Killing metric\footnote{The Cartan-Killing
metric $\eta^{ab}= \Tr T^a T^b$ is the natural metric on the group manifold, and its
positive-definiteness is equivalent to compactness for finite-dimensional algebras.}
 is positive definite and whose root lattice is bounded. (i.e.,
there is some real number greater than the magnitude of any root)
For finite-dimensional algebras, the root lattice is always bounded so pseudocompactness
is the same as compactness. A simple example of an algebra which satisfies the
first but not the second condition, to which we will refer below, is the Virasoro algebra.
From here on out we will be concerned only with countably
infinite-dimensional (``countable'') pseudocompact Lie algebras. We wish to prove

{\bf Proposition 1:} The unique (up to isomorphism) simple, countable pseudocompact
Lie algebra is $SU(\infty)$.

and

{\bf Proposition 2:} The Poisson algebra of any compact, connected  and orientable
2-manifold (i.e. a Riemann surface) is $U(1)$ times a simple, countable pseudocompact
Lie algebra.

First we will show that Dynkin diagrams are meaningful for pseudocompact algebras.
Dynkin diagrams are a way of graphically representing the simple roots of the algebras,
so what we must show is that the simple roots can be defined and that they contain
all of the information necessary to construct the algebra. We also need to find a few
properties of the simple roots which will define the conditions for a given lattice to
be a valid simple root lattice, e.g. linear independence.

We begin with the weights and roots. Any Lie algebra has an adjoint representation. For a
countable algebra this is represented on a separable Hilbert space, and for generators $T_a$
\be
T_a\ket{T_b} = \ket{T_c}\bra{T_c}T_a\ket{T_b} = -if^c_{ab}\ket{T_c} = \ket{[T_a,T_b]}\ .
\ee
So let $H$ be a maximal commuting subalgebra (the Cartan subalgebra) of the algebra.
Since the Cartan elements $H_m$ are self-adjoint and compact (this compactness
following from the finite norm of the roots), the Hilbert-Schmidt theorem guarantees
that their eigenvectors form a basis which simultaneously diagonalizes all of the
$H_m$'s. The eigenvalues are labelled by $\alpha$, (which by definition are the root
vectors) and the eigenkets $\ket{E_\alpha}$ (being a basis for the adjoint representation)
correspond to eigenoperators $E_\alpha$ of the Cartan subalgebra
\be
[H_m, E_\alpha]=\alpha_m E_\alpha\ .
\label{eq:ladderop}
\ee

It is straightforward that $E^\dagger_\alpha = E_{-\alpha}$, and thus roots
must occur in pairs. So to each root $\alpha$ there corresponds an $SU(2)$
subalgebra $E_+^{(\alpha)}=|\alpha|^{-1}E_\alpha,\ E_-^{(\alpha)} = |\alpha|^{-1}E_{-\alpha},\
E_3^{(\alpha)} = |\alpha|^{-2}\alpha\cdot H$. Since the roots all have finite norm these
are well-defined quantities.

Now let $\ket{\mu}$ be any state in a representation of the algebra. Its $E_3^{(\alpha)}$
eigenvalue is $\frac{\a\cdot\mu}{\a^2}$, so $\frac{2\a\cdot\mu}{\a^2}$ must be
an integer if the $SU(2)$ representation to which it belongs is finite. More
strongly, we can say that if it belongs to a finite-dimensional representation
of $SU(2)$, then
\be
\frac{2\a\cdot\mu}{\a^2} = q-p
\ee
where $p$ is the number of times $E_+^{(\alpha)}$ can act on $\mu$ without getting zero,
and $q$ is the number of times $E_-^{(\alpha)}$ can do the same. This equation follows
from the fact that $\a\cdot\mu/\a^2$ is the eigenvalue of $E_3^{(\alpha)}$ and the usual
properties of $SU(2)$ representations.

One particular case of interest is the adjoint representation of the algebra.
If one starts with any state $\ket{E_\b}$ and acts on it repeatedly with
$E_\pm^{(\alpha)}$, one gets a complete representation of the $SU(2)$ associated with $\alpha$.
This representation
would consist of states with roots $\b+n\a$ for integer $n$, and since the root lattice is
bounded there cannot be more than finitely many such roots. This means that
for any $\a$ and $\b$, $2\a\cdot\b/\a^2 = q-p$ for integers $q$ and $p$.

The next step is to define simple roots. We pick a sign convention for the roots
such that every nonzero root is either positive or negative, and the negative of
a positive root is a negative root. Then a simple root is defined to be a positive
root which is not the sum of any other positive roots. We claim that the simple
roots contain enough information to reconstruct the roots, and therefore the
algebra.

First, every positive root can be written as an integer linear combination of the
simple roots. To see this, consider some positive root $\g$. If $\g$ is simple, the
statement is trivial; otherwise $\g=\g_1+\g_2$ where $\g_1$ and $\g_2$ are positive.
We repeat this argument indefinitely, stopping whenever a root is simple. This must
terminate after no more than countably many steps, since every root has finite norm
and the coefficients in the sum are all nonnegative integers.

Next, the simple roots are linearly independent. This is because the most general
linear combination of simple roots can be written as a sum of terms with positive and
negative coefficients;
\be
\g = \sum_\a \mu_\a \a - \sum_\a \nu_\a \a
\ee
where the $\mu_\a$ and the $\nu_\a$ are all nonnegative integers, and for any given
$\a$ no more than one of $\mu_\a$ and $\nu_\a$ is nonzero. Then
\be
|\g|^2 = |\mu|^2 + |\nu|^2 - 2 \sum_{\a\not=\b}\mu_\a \nu_\b \a\cdot \b
\label{eq:gnorm}
\ee
However, $\a\cdot\b$ must be nonpositive. This can be shown as follows. First, for
any two distinct simple roots $\a$ and $\b$, $\a-\b$ is not a root, since if it
were then either $\a-\b>0$, in which case $\a=(\a-\b)+\b$, or $\b-\a>0$, in which
case $\b=(\b-\a)+\a$, violating the simplicity of $\a$ and $\b$ in either case.
This means that $E_{-\a}\ket{E_\b}\propto \ket{E_{\b-\a}}=0$, so $2\a\cdot\b/\a^2=-p\le 0$.
(Since the relevant $q$ is by definition zero) This means that the third term in
(\ref{eq:gnorm}) is nonnegative, so $|\g|^2\ge 0$. By positivity of the metric, this
means that it vanishes only for $\g=0$, so no (finite or infinite) linear combination
of the simple roots can vanish.

This therefore means that the simple roots form a basis for the set of all roots. Since
there are countably many roots and the root lattice is bounded, the root lattice has
countably infinite dimension, and so the number of simple roots is countably infinite
as well.

Now at last we can construct the roots from the simple roots. Any positive root can
be written as
\be
\phi = \sum k_\a \a
\ee
for positive integer coefficients $k_\a$. We define the level of the root to be the sum
of these coefficients, and find the roots by induction on the level. The set of level-1
roots is just the set of simple roots. Then given the set of level-$(\ell-1)$ roots, the
level-$\ell$ roots are given by
\be
\ket{E_{\phi_\ell}} = E_\a \ket{E_{\phi_{\ell-1}}}
\label{eq:levelmove}
\ee
for every $\a$ and $\phi_\ell$ for which the right-hand side is not zero. That can be
determined using
\be
\frac{2\a\cdot\phi_{\ell-1}}{\a^2} = q-p
\label{eq:peval}
\ee
where $p$ is the number of times $\ket{E_{\phi_{\ell-1}}}$ can be acted on by $E_\a$
without reaching zero. $\phi_{\ell-1}+\a$ is a root iff $p>0$. But $q$ is known since we
know how $\ket{E_{\phi_{\ell-1}}}$ was built out of the simple roots, so $p$ can be
evaluated from (\ref{eq:peval}) and (\ref{eq:levelmove}) generates the set of level-$\ell$
roots.

This can only fail if some $\ket{E_{\phi_\ell}}$ cannot be written in the form (\ref{eq:levelmove}).
The fact that this is impossible can be seen as follows. Knapp corollary 2.37 states
that whenever no multiple other than $1$, $0$ or $-1$ of any root is a root (which is
always true for finite-dimensional algebras; cf. Knapp proposition 2.21) the operator
$E_\a E_{-\a}$ is nondegenerate for all $\alpha$. Pseudocompact algebras must satisfy this condition,
since for any root $\a$ the set $\{\a\cdot H, E_{n\a}\}$ for all integers $n$ for which
$n\a$ is a root is a subalgebra of our algebra, and by boundedness of the root lattice it
is finite-dimensional. Then the usual argument that in a finite-dimensional algebra no
multiple of a root greater than 1 is a root applies, and so $n\a$ cannot be a root for
$|n|>1$.

This tells us that $\ket{E_{\phi_\ell}}$ is annihilated
by every $E_{-\a}$, since if it were not then $E_{-\a}\ket{E_{\phi_\ell}}$ would be a
level-$(\ell-1)$ root and acting on it with $E_\a$ would give $\ket{E_{\phi_\ell}}$ in
the form (\ref{eq:levelmove}).\footnote{For algebras which are not pseudocompact, this
argument does not hold. For example, the Virasoro algebra $[L_m, L_n]=i(m+n)L_{m+n}$
has one element, $L_0$, in its Cartan subalgebra, and root vectors $E_n=L_{-n}$.
Its root lattice is unbounded, and so for example $E_1E_{-1}$ is degenerate;
$L_{-1}L_1\ket{L_{-2}}=\ket{[L_{-1}, [L_1, L_{-2}]]}=0$. As a result the usual root-building
algorithm fails to find the positive root $L_{-2}$ and so this algebra cannot be reconstructed
from its simple roots alone. This is why boundedness of the root lattice is so important.}
(It would not be zero since $E_\a E_{-\a}$ is nondegenerate)
Therefore for each simple root $\a$, $E_{-\a}\ket{E_{\phi_\ell}}=0$ and so from
our usual formula $2\a\cdot\phi_\ell/\a^2=-p\le 0$. This means that $\phi_\ell^2=
\sum k_\a \a\cdot\phi_\ell\le 0$ and (by positive definiteness) $\phi_\ell=0$, a
contradiction. Thus every positive root can be written in the form (\ref{eq:levelmove})
and so is reached by this inductive process.

Therefore the simple roots carry all of the information needed to reconstruct the algebra,
and thus there is a one-to-one correspondence between simple root lattices and countable
pseudocompact Lie algebras. The simple roots must be linearly independent, the angles
between them must satisfy $\cos\theta_{\a\b}=-\sqrt{pp'}/2 = 0,\ 1/2,\ 1/\sqrt{2},$ or
$\sqrt{3}/2$, and their length ratios must satisfy $|\a|^2/|\b|^2=p/p'$. We can therefore
introduce Dynkin diagrams to represent the simple root lattices, drawing a circle for each
simple root and joining the circles $\a$ and $\b$ by $4\cos^2\theta_{\a\b}$ lines. The
set of countable pseudocompact Lie algebras is isomorphic to the set of Dynkin diagrams
with countably many circles.

We can now repeat the ordinary classification of Dynkin diagrams. Simple algebras correspond
to connected diagrams since if the simple roots decompose into two orthogonal subsets (and thus
two networks with $\cos^2\theta_{\a\a'}=0$ between elements of the two subsets) the
corresponding ladder operators commute and thus the algebra factors.

Linear independence implies the following five standard lemmas:
(The proofs are given in Georgi chapter 20 or Knapp chapter 2)

{\bf Lemma 0.} Every subdiagram of a Dynkin diagram is itself a consistent Dynkin diagram.

{\bf Lemma 1.} The only Dynkin diagrams with three root vectors
are those given in figure \ref{fig:lem1configs}.

{\bf Lemma 2.} If a diagram contains two vectors connected by a
single line, the diagram obtained by shrinking the line away and merging the
two vectors into a single circle represents another consistent diagram.

{\bf Lemma 3.} In figure \ref{fig:lem3configs}, if the first configuration is
a consistent diagram, then the second is as well.

\begin{figure}[tbhp]
\centerline{\epsffile{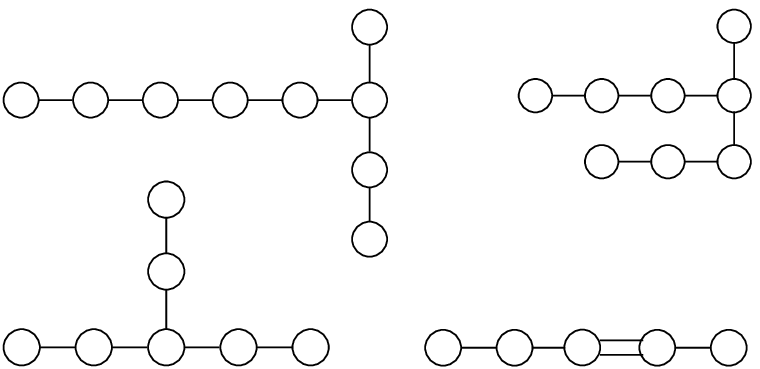}}
\caption{Lemma 4}
\label{fig:lem4configs}
\end{figure}

{\bf Lemma 4.} No diagram contains any of the subdiagrams in figure \ref{fig:lem4configs}.

\begin{minipage}[tbhp]{0.4\textwidth}
\begin{figure}[bthp]
\centerline{\epsffile{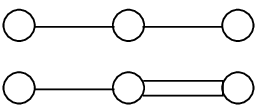}}
\caption{Lemma 1}
\label{fig:lem1configs}
\end{figure}
\end{minipage}
\begin{minipage}[thbp]{0.4\textwidth}
\begin{figure}[bthp]
\centerline{\epsffile{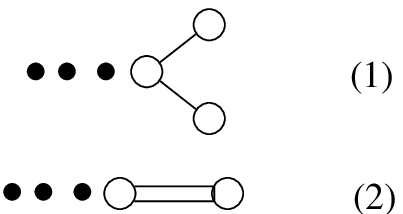}}
\caption{Lemma 3}
\label{fig:lem3configs}
\end{figure}
\end{minipage}

 These do not rule out some
nontrivial infinite diagrams, since lemma 2 only allows us to contract finite chains of
singly-connected roots. (There may be subtleties for infinite chains) We therefore need to
prove the additional lemma

\begin{figure}[tbhp]
\centerline{\epsfxsize=0.8\textwidth  \epsffile{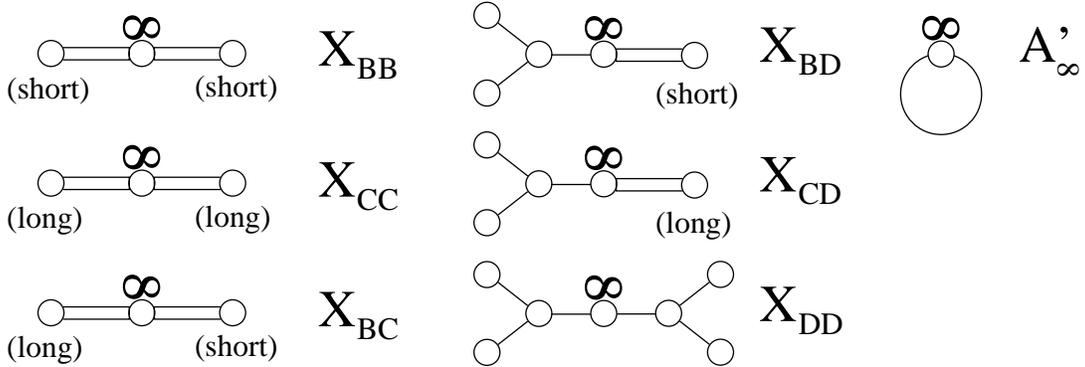}}
\caption{Lemma 5. A circle with an $\infty$ symbol represents an infinite chain of
circles connected by single lines; $A'_\infty$ is an infinite loop.}
\label{fig:xalgebras}
\end{figure}

{\bf Lemma 5.} The diagrams shown in figure \ref{fig:xalgebras} are not consistent.

We give an explicit proof for $X_{BB}$; the remaining cases are
essentially identical. If we
had a rank-$(n+2)$ version of this algebra, with the infinite chain replaced
by a chain of length $n$, then from the diagram we can read
\be
\alpha_0^2 = \alpha_{n+1}^2 = 1;\ \alpha_i^2 = 2
\ee
where $i=1\ldots n$, and
\be
\alpha_i\cdot \alpha_{i\pm 1} = \alpha_0\cdot\alpha_1 = \alpha_n
\cdot\alpha_{n+1} =-1
\ee
with all other products vanishing. Therefore the linear combination
$\sum\alpha$ is orthogonal to every root, since
\bea
(\sum\alpha)\cdot \alpha_1 &=& \alpha_1^2 + \alpha_1\cdot\alpha_2 = 0 \nn \\
(\sum\alpha)\cdot \alpha_2 &=& \alpha_2^2 + \alpha_1\cdot\alpha_2 + \alpha_3
\cdot\alpha_2 = 0
\eea
etc. Since the simple roots are a basis, this means that $\sum\alpha=0$ and thus these
roots fail to be linearly independent. Since
this can be done for any $n$, this combination is an obvious ansatz
for the infinite case $X_{BB}$. There is no subtlety in the limit
$n\rightarrow\infty$ because the criterion for linear independence of the simple
roots does not rely on the finiteness of the linear combination. A similar calculation
follows for each of the remaining algebras.

\begin{figure}[hbtp]
\centerline{\epsffile{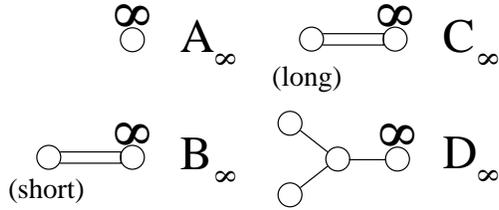}}
\caption{The infinite classical algebras}
\label{fig:classical}
\end{figure}

Therefore the only consistent pseudocompact Lie algebras are the four infinite
classical algebras. (Figure \ref{fig:classical})
We will now show that these four algebras are actually identical. (The
following can also be seen explicitly by writing the various algebras in
a basis of clock and shift matrices; a complete discussion has been given
by Fairlie {\em et al.} \cite{Fairlie}) We need for this a standard lemma about algebras:

\medskip

{\bf Lemma.} Let $U_n$ and $V_n$ be two sequences of algebras which have
inductive limits $U$ and $V$, respectively. If for each $n$, $U_n\subset
V_{n+1}$ and $V_n\subset U_n$, then $U\cong V$.\footnote{A simple proof of
this using commutative diagrams is given in lemma L.1.5 of Wegge-Olsen.\cite{WeggeOlsen}}

\medskip

The simplest application of this lemma is to show that $B_\infty=D_\infty$.
These two algebras are the inductive limits of the sequences $B_n=SO(2n+1)$
and $D_n=SO(2n)$, respectively. Clearly $SO(2n)\subset SO(2n+1)$ and
$SO(2n+1)\subset SO(2n+2)$, and therefore the two limits must be equal. This is
just the statement that $\lim_{n\rightarrow\infty} SO(n)$ is the same
whether one progresses by even or odd $n$.

Similarly, we can show that $B_\infty=A_\infty$. For every $n$, $SO(n)\subset SU(n)$
since every orthogonal matrix is unitary. (This is because every
real number is also a complex number) It is also well-known (and clear from
Dynkin diagrams) that for each $n$, $SU(n)\subset SO(2n)$. Thus using the sequences
$U_n=SU(2^n)$ and $V_n=SO(2^n)$, the claim is proved.

Finally, we can show that $A_\infty=C_\infty$. For every $n$, $SU(n)\subset Sp(2n)$,
since every complex number is also a quaternion. Similarly $Sp(2n)\subset SU(2n)$
since the generators of each $Sp$ group are Hermitian and therefore are also
generators of the $SU$ group of the same dimension.
Therefore all four of the classical infinite algebras are isomorphic, and so we have
proven the first proposition.

\bigskip

Now we turn to the second proposition. Since our manifold is orientable,
it automatically has a Poisson structure defined by setting its symplectic form
equal to its volume form. The dimension of the Poisson algebra is countably infinite
by Fourier's theorem; namely, the set of functions on a compact manifold has a
countable basis.\footnote{
The contrapositive of this theorem gives us an immediate and important result,
alluded to above, that if the membrane were unbounded (e.g. as the plane) then its Poisson algebra would
have uncountably infinite dimension. The remainder of the proof below will still
apply, telling us that the Poisson algebra is still $U(1)$ times a simple pseudocompact
algebra, but we do not have any useful results on the classification of these much
larger algebras. But these algebras clearly cannot be approximated by matrix algebras,
since a sequence of finite-dimensional objects cannot converge to one of uncountable
dimension.} The elements of the Poisson algebra must be smooth functions, since
functions of finite differentiability class would not be closed under the Poisson
bracket.

We can see that the algebra has a $U(1)$ factor since the set of constant functions is
1-dimensional and has vanishing Poisson bracket with all functions. That the algebra
is otherwise simple can be seen as follows. There is a one-to-one correspondence between
invariant subalgebras of the Poisson algebra $P$ and projections on the algebra, that is
to say mappings $\phi:P\rightarrow P$ which preserve the Lie bracket and satisfy $\phi^2=\phi$.
Explicitly, for any invariant subalgebra $H$, the mapping $\phi_H(p)=p$ if $p\in H$ and
0 if $p\not\in H$ is a projection; conversely, for any projection $\phi$ and $a,b\in P$ such that
$\phi(a)=a$ and $\phi(b)=0$, (the only eigenvalues of a projection are zero and one, so every
element of $P$ satisfies one or the other condition) then
$\phi([a,b])=[\phi(a),\phi(b)]=0$, and $(1-\phi)([a,b])=[(1-\phi)(a), (1-\phi)(b)]=0$,
so $[a,b]=0$ and thus the algebra factors into $P=\phi(P)\oplus (1-\phi)(P)$. Projections
on the Poisson algebra are clearly also projections on the algebra of smooth functions under
multiplication, and vice versa. (After all, both are algebras of functions; only their product
rule is different) A projection in the algebra of smooth functions is a function which is
everywhere either zero or one, and so is piecewise constant, and so the number of invariant
subalgebras of $P$ is equal to the number of piecewise constant smooth functions on the
manifold, that is to say the number of components. For a connected manifold, which is the
case of interest to us, the only piecewise constant function is the constant function,
corresponding to the invariant $U(1)$ subalgebra noted above.

Finally, we must prove that the algebra is pseudocompact, i.e. that its Cartan-Killing
metric is positive definite and that its root lattice is bounded. This is straightforward
since it is easy to show that the Poisson algebra must be a subgroup of $U(\infty)$,
and any subalgebra of a pseudocompact algebra must be pseudocompact as well. To see this,
we note that for any real-valued function, the generator associated to it is
\be
T_f = i\left\{f, \cdot\right\} = i\Omega^{ij}\partial_i f \partial_j
\ee
which is Hermitian. Since all the generators are Hermitian and the algebra is countable,
the algebra must be a subalgebra of the algebra of all Hermitian, countable matrices,
which is precisely $U(\infty)$. This completes the proof of proposition 2, which together
with proposition 1 proves our theorem.

It is interesting that the proof of proposition 2 did not require that the manifold be
2-dimensional at any point; in fact, it applies equally well to any compact symplectic manifold. These
manifolds are not the phase spaces of ordinary systems, since momentum is generally not
constrained, but they may be natural phase spaces for systems which exhibit UV/IR dualities
of various sorts, including T-duality.

\medskip

We have therefore shown that there is a systematic matrix approximation to the theory of
compact relativistic orientable membranes, in the sense that there is an infinite sequence of
classical Lagrangians which converges in the large-$N$ limit to the Lagrangian
(\ref{eq:superbrane1}). We have also seen that membrane theories which do not satisfy
these constraints -- theories of infinite branes or theories which allow nonorientable
branes -- do {\em not} seem to have a meaningful matrix approximation, or at least not
one by dimensionally reduced Yang-Mills theories. This latter conclusion may be interesting
from the perspective of the bosonic M-theory recently proposed by Horowitz and Susskind.
\cite{BM27} Applied to traditional M-theory, this is further evidence for the hypothesis
that the matrix model carries the full structure of the theory.

\medskip

The author would like to thank Savas Dimopoulos, Simeon Hellerman, Matthew Kleban,
Steve Shenker and Leonard Susskind for helpful comments. This work was supported by the
National Science Foundation under grant number PHY-9870015.

\end{document}